\documentclass{svjour3}

\usepackage{amsmath,amssymb}
\usepackage{hyperref}
\usepackage{subfig}

\usepackage[amsmath, thmmarks]{ntheorem}
{\theoremstyle{nonumberplain}
\theoremsymbol{\ensuremath{\Box}}
\newtheorem{pf}{Proof}}

\newtheorem{thm}{Theorem}
\newtheorem{lem}{Lemma}

\newtheorem{ex}{Example}

\title{Two-weight and three-weight linear codes based on Weil sums}

\author{Gaopeng Jian}    

\institute{Gaopeng Jian \at Key Laboratory of Machine Perception(MOE), School of EECS, Peking University, Beijing 100871, P.R.China; \\
\email{gpjian@pku.edu.cn}
}

\date{}

\begin{document}

\maketitle

\begin{abstract}
Linear codes with few weights have applications in secret sharing, authentication codes, association schemes and strongly regular graphs.
In this paper, several classes of two-weight and three-weight linear codes are presented  and their weight distributions are determined using Weil sums. 
Some of the linear codes obtained are optimal or almost optimal with respect to the Griesmer bound.
\end{abstract}

\section{Introduction}
Let $\mathbb{F}_p$ be the finite field with $p$ elements, where $p$ is an odd prime.
An $[n,k,d]$ linear code $C$ over $\mathbb{F}_p$ is a $k$-dimensional subspace of $\mathbb{F}_p^n$ with minimum Hamming distance $d$.
Let $A_i$ be the number of codewords with Hamming weight $i$ in $C$. 
The polynomial $1+A_1 z+A_2 z^2+\cdots+A_n z^n$ is called the weight enumerator of $C$ and the sequence $(1, A_1, A_2, \ldots, A_n)$ called the weight distribution of $C$.
If the number of nonzero $A_i$ in the sequence $(A_1, A_2, \ldots, A_n)$ is equal to $t$, we call $C$ a $t$-weight code.
The weight distribution of a code contains important information on its error correcting capability and the error probability of its error detection and correction with respect to some algorithms \cite{torleiv2007codes}.
In addition, much attention has been paid to two-weight and three-weight linear codes  \cite{tang2016linear,zhou2015linear,heng2016three,heng2016two,heng2015class,heng2017construction,luo2018binary,ding2014binary,ding2015class} due to their applications in secret sharing \cite{yuan2006secret,carlet2005linear}, strongly regular graphs \cite{calderbank1986geometry}, association schemes \cite{calderbank1984three} and authentication codes \cite{ding2005coding}. 

Let $\text{Tr}$ denote the trace function from $\mathbb{F}_q$ to $\mathbb{F}_p$.
For a set $D=\{d_1,d_2,\ldots,d_n \}\subset \mathbb{F}_q$, define a $p$-ary linear code of length $n$ by
\[
C_D=\{c(a)=(\text{Tr}(a d_1), \text{Tr}(a d_2), \ldots, \text{Tr}(a d_n)): a \in \mathbb{F}_q\}.
\]
The set $D$ is called the \emph{defining set} of $C_D$.
The construction was first proposed by Ding \emph{et al.} in \cite{ding2007cyclotomic} and many classes of known codes could be produced by properly selecting the defining set. 

Motivated by the above construction, Li \emph{et al.} \cite{li2016construction} defined a $p$-ary linear code by
\begin{equation}
C_D=\{c(a,b)=(\text{Tr}(ax+by))_{(x,y) \in D}: a,b \in \mathbb{F}_q \}
\label{cdd}
\end{equation}
where $D \subset \mathbb{F}_q^2$ is also called a defining set. 
It can be viewed as a generalization of cyclic codes whose duals have two zeros. 
In this paper, we extend their results by choosing different defining sets given by
\begin{equation} \label{d1}
D_1=\{ (x,y) \in \mathbb{F}_q^2 \backslash \{(0,0)\}: \text{Tr}(x+y^{p^u+1})=0\}
\end{equation}
and 
\begin{equation} \label{d2}
D_2=\{ (x,y) \in \mathbb{F}_q^2 \backslash \{(0,0)\}: \text{Tr}(x^2+y^{p^u+1})=0\}.
\end{equation}
We mainly use Weil sums to determine their parameters and weight distributions.
In particular, some of the linear codes obtained are optimal or almost optimal with respect to the Griesmer bound \cite{griesmer1960bound}.

The rest of this paper is organized as follows. 
In section \ref{pl} we introduce some basic notations and results on group characters, character sums and Pless power moments, which will be employed  later.
In section \ref{state} we present the parameters of several classes of two-weight and three-weight linear codes.  We also give some examples.
Section \ref{wt} is devoted to giving the proofs of main results. 
Section \ref{clu} summarizes this paper.

\section{Preliminaries} \label{pl}
From now on we fix the following notations.
\begin{itemize}
\item $q=p^m$, where $p$ is an odd prime and $m$ is a positive integer.
\item $s=\frac{m}{2}$.
\item $u$ is a positive integer, $v=gcd(m,u)$.
\item $\text{Tr}$ is the trace function from $\mathbb{F}_q$ to $\mathbb{F}_p$.
\item $\zeta_p=e^{\frac{2 \pi \sqrt{-1}}{p}}$ is the primitive $p$-th root of unity.
\item $\eta$ and $\eta_p$ are the quadratic multiplicative characters of $\mathbb{F}_q$ and $\mathbb{F}_p$, respectively
\end{itemize}

\subsection{Group characters and Gauss sums}
An additive character of $\mathbb{F}_q$ is a homomorphism from the additive group $\mathbb{F}_q$ into the multiplicative group composed by the $p$-th roots of unity in the complex numbers. 
For each $b \in \mathbb{F}_q$, the function
\[
\chi_b(x) = \zeta_p^{\text{Tr}(bx)} \text{ for all $x \in \mathbb{F}_q$}
\]
defines an additive character of $\mathbb{F}_q$. 
It is clear that $\chi_0 (x) = 1$ for all $x \in \mathbb{F}_q$ and it is called the trivial additive character of $\mathbb{F}_q$. 
The character $\chi:=\chi_1$ is called the canonical additive character of $\mathbb{F}_q$. 
It is easy to see that $\chi_b(x)=\chi(bx)$ for all $b,x \in \mathbb{F}_q$. 
The orthogonal property of additive characters is given by
\[
\sum_{x \in \mathbb{F}_q} \chi_b(x)=\begin{cases}
q, & \text{if} \ b=0, \\
0, & \text{otherwise}.
\end{cases}
\]

A multiplicative character of $\mathbb{F}_q$ is a homomorphism from the multiplicative group $\mathbb{F}_q^*=\mathbb{F}_q \backslash \{0\}$ into the multiplicative group composed by the $(q-1)$-th roots of unity in the complex numbers.
Let $\lambda$ be a multiplicative character of  $\mathbb{F}_q$, we define the Gauss sum over $\mathbb{F}_q$ by
\[
G(\lambda)=\sum_{x \in \mathbb{F}_q^*}\lambda(x) \chi(x).
\]

The explicit value of Gauss sums are very difficult to determine and are known for only a few cases.
For future use, we state some results about Gauss sums. The quadratic Gauss sums are known and given in the following lemma.

\begin{lem}[{{\cite[Theorem 5.15]{lidl1997finite}}}] \label{ga}
\[
G(\eta)=(-1)^{m-1}\sqrt{(p^*)^m}=\begin{cases}
(-1)^{m-1} \sqrt{q}, & \text{if} \ p \equiv 1 \pmod{4},\\
(-1)^{m-1}(\sqrt{-1})^m \sqrt{q}, &  \text{if} \ p \equiv 3 \pmod{4},
\end{cases}
\]
where $p^*=(-1)^{\frac{p-1}{2}}p$.
\end{lem}

\begin{lem}[{{\cite[Lemma 7]{ding2015class}}}] \label{e}
For $x \in \mathbb{F}_p^*$,
\[
\eta(x)=\begin{cases}
1, & \text{if $m$ is even}, \\
\eta_p(x),  & \text{if $m$ is odd}. \\
\end{cases}
\]
\end{lem}

\subsection{Weil sums}
Weil sums are defined by $\sum_{x \in \mathbb{F}_q}\chi(f(x))$ where $f(x) \in \mathbb{F}_q[X]$.
In \cite{coulter1998explicit,coulter1998further}, Coulter evaluated some Weil sums given by
\[
S_u(a,b)=\sum_{x \in \mathbb{F}_q}\chi(ax^{p^u+1}+bx), \ a \in \mathbb{F}_q^*, \ b \in \mathbb{F}_q.
\]

\begin{lem} \label{s3}
If $\frac{m}{v}$ be odd, then
\[
S_u(a,0)=G(\eta) \eta(a)=\begin{cases}
(-1)^{m-1} \sqrt{q}\eta(a), & \text{if} \ p \equiv 1 \pmod{4},\\
(-1)^{m-1}(\sqrt{-1})^m \sqrt{q}\eta(a), &  \text{if} \ p \equiv 3 \pmod{4}.
\end{cases}
\]
\end{lem}

\begin{lem}  \label{s4}
If $\frac{m}{v}$ is even, then
\[
S_u(a,0)=\begin{cases}
p^s, & \text{if $\frac{s}{v}$ is even and } a^{\frac{q-1}{p^v+1}} \ne (-1)^{\frac{s}{v}},\\
-p^{s+v}, & \text{if $\frac{s}{v}$ is even and } a^{\frac{q-1}{p^v+1}} = (-1)^{\frac{s}{v}},\\
-p^s, & \text{if $\frac{s}{v}$ is odd and } a^{\frac{q-1}{p^v+1}} \ne (-1)^{\frac{s}{v}},\\
p^{s+v}, & \text{if $\frac{s}{v}$ is odd and } a^{\frac{q-1}{p^v+1}} = (-1)^{\frac{s}{v}}.\\
\end{cases}
\]
\end{lem}

\begin{lem} \label{sol}
The equation 
\[
a^{p^u} X^{p^{2u}}+aX=0
\]
is solvable over $\mathbb{F}_q^*$ if and only if $\frac{m}{v}$ is even and $a^{\frac{q-1}{p^v+1}}=(-1)^{\frac{s}{v}}$.
In such cases there are $p^{2v}-1$ non-zero solutions.
\end{lem}

\begin{lem} \label{s1}
Suppose $f(X) = a^{p^u}X^{p^{2u}}+aX$ is
a permutation polynomial over $\mathbb{F}_q$. 
Let $x_0$ be the unique solution of the equation $f(X)=-b^{p^u}$. The evaluation of $S_u(a,b)$ partitions into the following two cases:
\begin{enumerate}
\item If $\frac{m}{v}$ is odd, then
\begin{align*}
S_u(a,b)&=G(\eta) \eta(a) \bar{\chi} \left( ax_0^{p^u+1}\right) \\
&=\begin{cases}
(-1)^{m-1} \sqrt{q}\eta(a) \bar{\chi} \left( ax_0^{p^u+1}\right), & \text{if} \ p \equiv 1 \pmod{4},\\
(-1)^{m-1}(\sqrt{-1})^m \sqrt{q}\eta(a) \bar{\chi} \left( ax_0^{p^u+1}\right), &  \text{if} \ p \equiv 3 \pmod{4}.
\end{cases}
\end{align*}
\item If $\frac{m}{v}$ is even, then $a^{\frac{q-1}{p^v+1}} \ne (-1)^{\frac{s}{v}}$ and
\[
S_u(a,b)=(-1)^{\frac{s}{v}}p^s \bar{\chi} \left( ax_0^{p^u+1}\right).
 \] 
\end{enumerate}
\end{lem}

\begin{lem} \label{s2}
Suppose $f(X) = a^{p^u}X^{p^{2u}}+aX$ is not a permutation polynomial over $\mathbb{F}_q$,  then $S_u(a,b)=0$ unless the equation $f(X) = -b^{p^u}$ is solvable. If the equation is solvable, with some solution $x_0$ say, then 
\[
S_u(a,b)=-(-1)^{\frac{s}{v}}p^{s+v} \bar{\chi} \left( ax_0^{p^u+1}\right).
 \] 
\end{lem}

The following lemma is a special case ($u=0$) of Lemma \ref{s3} and \ref{s1} and is also proved in \cite[Theorem 5.33]{lidl1997finite}.

\begin{lem} \label{qu}
\[
Q(a,b)=\sum_{x \in \mathbb{F}_q}\chi(ax^2+bx)
=G(\eta) \eta(a) \bar{\chi} \left(\frac{b^2}{4a}\right),\ a \in \mathbb{F}_q^*, \ b \in \mathbb{F}_q.
\]
\end{lem}

\subsection{The Pless power moments}
For an $[n,k,d]$ code $C$ over $\mathbb{F}_p$ with weight distribution $(1,A_1,\ldots,A_n)$, we denote by
$(1, A_1^{\perp},\ldots, A_n^{\perp})$ the weight distribution of its dual code.
The first two Pless power moments are given as follows \cite[p.259]{huffman2010fundamentals}:
\[
\sum_{j=0}^n A_j=p^k,
\]
\[
\sum_{j=0}^n jA_j=p^{k-1}(pn-n-A_1^{\perp}).
\]
For the code $C_D$ defined by \eqref{cdd}, $A_1^{\perp}=0$ if $(0,0) \notin D$ by the nondegenerate property of the trace function.

\section{The statements of main results} \label{state}
\begin{thm} \label{th1}
Let $C_{D_1}$ be defined by \eqref{cdd}, where $D_1$ is defined in \eqref{d1}. 
If $m$ is odd, then $C_{D_1}$ is a $[p^{2m-1}-1, 2m]$ three-weight linear code with weight distribution given in Table \ref{t2}.
\end{thm}

\begin{table} 
\caption{The weight distribution of $C_{D_1}$ if $m$ is odd}
\centering
\begin{tabular}{ll} 
\hline
Weight $w$                & Multiplicity $A_w$ \\
\hline
0 &1 \\
$(p-1)p^{2m-2}$ &  $p^{2m}-1-(p-1)^2 p^{m-1}$\\
$(p-1)p^{2m-2}\left( 1-\frac{1}{(p-1)p^{\frac{m-1}{2}}} \right)$ & $\frac{1}{2}(p-1)^2\left( p^{m-1}+p^{\frac{m-1}{2}} \right)$\\
$(p-1)p^{2m-2}\left( 1+\frac{1}{(p-1)p^{\frac{m-1}{2}}} \right)$ & $\frac{1}{2}(p-1)^2\left( p^{m-1}-p^{\frac{m-1}{2}} \right)$ \\
\hline
\end{tabular}
 \label{t2}
\end{table}

\begin{ex}
Let $p=3$ and $m=3$. 
Then the code $C_{D_1}$ has parameters [242,6,135] and weight enumerator $1 + 24z^{135} + 692z^{162} + 12z^{189}$. 
\end{ex}

\begin{thm} \label{th2}
Let $C_{D_1}$ be defined by \eqref{cdd}, where $D_1$ is defined in \eqref{d1}.  
If $\frac{m}{v}$ is odd and $v$ is even, then $C_{D_1}$ is a $[p^{2m-1}-1, 2m]$ three-weight linear code  with weight distribution given in Table \ref{t1}.
\end{thm}

\begin{table} 
\caption{The weight distribution of $C_{D_1}$ if $\frac{m}{v}$ is odd and $v$ is even}
\centering
\begin{tabular}{ll} 
\hline
Weight $w$                & Multiplicity $A_w$ \\
\hline
0 &1 \\
$(p-1)p^{2m-2}$ & $p^{2m}-1-(p-1)p^m$ \\
$(p-1)p^{2m-2}\left( 1-\frac{G(\eta)}{q} \right)$ & $(p-1)p^{m-1}\left(1+\frac{(p-1)G(\eta)}{q}\right)$ \\
$(p-1)p^{2m-2}\left( 1+\frac{G(\eta)}{(p-1)q} \right)$ & $(p-1)p^{m-1}\left(p-1-\frac{(p-1)G(\eta)}{q}\right)$  \\
\hline
\end{tabular}
 \label{t1}
\end{table}

\begin{ex}
Let $p=3$, $m=2$ and $u=4$. 
Then $v=2$, $\frac{m}{v}=1$ and $G(\eta)=3$.  
The code $C_{D_1}$ has parameters [26,4,12] and weight enumerator $1 + 10z^{12} + 62z^{18} + 8z^{21}$. 
\end{ex}

\begin{thm} \label{th3}
Let $C_{D_1}$ be defined by \eqref{cdd}, where $D_1$ is defined in \eqref{d1}.
If $\frac{m}{v} \equiv 2 \pmod{4}$, then $C_{D_1}$ is a $[p^{2m-1}-1, 2m]$ three-weight linear code  with weight distribution given in Table \ref{t3}.
\end{thm}

\begin{table}  
\caption{The weight distribution of $C_{D_1}$ if $\frac{m}{v} \equiv 2 \pmod{4}$}
\centering
\begin{tabular}{ll} 
\hline
Weight $w$                & Multiplicity $A_w$ \\
\hline
0 &1 \\
$(p-1)p^{2m-2}$ & $p^{2m}-1-(p-1)p^m$ \\
$(p-1)p^{2m-2}\left( 1+\frac{1}{p^s} \right)$ &  $(p-1)(p^{m-1}-p^s+p^{s-1})$\\
$(p-1)p^{2m-2}\left( 1-\frac{1}{(p-1)p^s} \right)$ & $(p-1)(p^s+1)(p^s-p^{s-1})$  \\
\hline
\end{tabular}
\label{t3}
\end{table}

\begin{ex}
Let $p=3$, $m=2$ and $u=3$. 
Then $v=1$, $\frac{m}{v}=2$ and $s=1$.
The code $C_{D_1}$ has parameters [26,4,15] and weight enumerator $1 + 16z^{15} + 62z^{18} + 2z^{24}$. 
\end{ex}

\begin{thm} \label{th4}
Let $C_{D_1}$ be defined by \eqref{cdd}, where $D_1$ is defined in \eqref{d1}.
If $\frac{m}{v} \equiv 0 \pmod{4}$, then $C_{D_1}$ is a $[p^{2m-1}-1, 2m]$ three-weight linear code  with weight distribution given in Table \ref{t4}.
\end{thm}

\begin{table} 
\caption{The weight distribution of $C_{D_1}$ if $\frac{m}{v} \equiv 0 \pmod{4}$}
\centering
\begin{tabular}{ll} 
\hline
Weight $w$                & Multiplicity $A_w$ \\
\hline
0 &1 \\
$(p-1)p^{2m-2}$ & $p^{2m}-1-(p-1)p^{m-2v}$ \\
$(p-1)p^{2m-2}\left( 1+\frac{1}{p^{s-v}} \right)$ &  $(p-1)(p^{m-2v-1}-p^{s-v}+p^{s-v-1})$\\
$(p-1)p^{2m-2}\left( 1-\frac{1}{(p-1)p^{s-v}} \right)$ &  $(p-1)(p^{s-v}+1)(p^{s-v}-p^{s-v-1})$ \\
\hline
\end{tabular}
 \label{t4}
\end{table}

\begin{ex}
Let $p=3$, $m=4$ and $u=3$. 
Then $v=1$, $\frac{m}{v}=4$ and $s=2$.
The code $C_{D_1}$ has parameters [2186,8,1215] and weight enumerator $1 + 16z^{1215} + 6542z^{1458} + 2z^{1944}$.
\end{ex}

\begin{thm} \label{th5}
Let $C_{D_2}$ be defined by \eqref{cdd}, where $D_2$ is defined in \eqref{d2}.
If $\frac{m}{v}$ is odd or $\frac{m}{v} \equiv 2 \pmod{4}$, then $C_{D_2}$ is an $[n, 2m]$ two-weight linear code with weight distribution given in Table \ref{tt2}, where 
\[
n=\begin{cases}
(p^m+1)(p^{m-1}-1), &  \text{if $p \equiv 3 \pmod{4}$ and $v$ is odd}, \\
(p^m-1)(p^{m-1}+1), & \text{otherwise}. 
\end{cases} 
\]
\end{thm}

\begin{table}   
\centering
\caption{The weight distribution of $C_{D_2}$ if $\frac{m}{v}$ is odd or $\frac{m}{v} \equiv 2 \pmod{4}$}

\subfloat[$p \equiv 3 \pmod{4}$ and $v$ is odd]{
\begin{tabular}{ll} 
\hline
Weight $w$                & Multiplicity $A_w$ \\
\hline
0 &1 \\
$(p-1)p^{2m-2}$ & $(p^m+1)(p^{m-1}-1)$ \\
$(p-1)p^{2m-2}(1-\frac{1}{p^{m-1}})$ & $(p^m+1)p^{m-1}(p-1)$ \\
\hline
\end{tabular}
}

\subfloat[otherwise]{
\begin{tabular}{ll} 
\hline
Weight $w$                & Multiplicity $A_w$ \\
\hline
0 &1 \\
$(p-1)p^{2m-2}$ & $(p^m-1)(p^{m-1}+1)$ \\
$(p-1)p^{2m-2}(1+\frac{1}{p^{m-1}})$ & $(p^m-1)p^{m-1}(p-1)$ \\
\hline
\end{tabular}
}

\label{tt2}
\end{table}

Note that for any $a \in \mathbb{F}_p^*$, $\text{Tr}((ax)^2+(ay)^{p^u+1})=a^2 \text{Tr}(x^2+y^{p^u+1})$.
Then we can select a subset $\overline{D_2}$ of $D_2$ such that $\bigcup_{a \in \mathbb{F}_p^*} a \overline{D_2}$ is a partition of $D_2$.
Hence, the code $C_{D_2}$ can be punctured into a shorter linear codes $C_{\overline{D_2}}$ whose weights can be obtained from the original $C_{D_2}$ by dividing the common divisor $p-1$. 

\begin{ex}
Let $p=3$ and $m=3$. 
Then the code $C_{D_2}$ has parameters [224,6,144] and weight enumerator $1 + 504z^{144} + 224z^{162}$.
The code $C_{\overline{D_2}}$ has parameters [112,6,72] and weight enumerator $1 + 504z^{72} + 224z^{81}$, which is almost optimal as the best linear code of length 112 and dimension 6 over $\mathbb{F}_3$ has minimum weight 73 according to the Griesmer bound.  
\end{ex}

\begin{ex}
Let $p=3$, $m=2$ and $u=3$.
Then $v=1$ and $\frac{m}{v}=2$.
The code $C_{D_2}$ has parameters [20,4,12] and weight enumerator $1 + 60z^{12} + 20z^{18}$.
The code $C_{\overline{D_2}}$ has parameters [10,4,6] and weight enumerator $1 + 60z^{6} + 20z^{9}$.
Both codes are optimal according to the Griesmer bound, and $C_{D_2}$ is different from the best known linear codes from the Magma BKLC(GF(3),20,4) which has a different weight enumerator $1 + 60z^{12} + 18z^{15}+2z^{18}$.
\end{ex}

\begin{ex}
Let $p=3$, $m=2$ and $u=4$.
Then $v=2$ and $\frac{m}{v}=1$.
The code $C_{D_2}$ has parameters [32,4,18] and weight enumerator $1 + 32z^{18} + 48z^{24}$.
The code $C_{\overline{D_2}}$ has parameters [16,4,9] and weight enumerator $1 + 32z^{9} + 48z^{12}$, which is optimal according to the Griesmer bound.
In addition, $C_{\overline{D_2}}$ is different from the best known linear codes from the Magma BKLC(GF(3),16,4) which has a different weight enumerator $1 + 50z^{9} + 30z^{12}$.
\end{ex}

\begin{ex}
Let $p=3$, $m=4$ and $u=2$. 
Then $v=2$ and $\frac{m}{v}=2$.
The code $C_{D_2}$ has parameters [2240,8,1458] and weight enumerator $1 + 2240z^{1458} + 4320z^{1512}$. 
The code $C_{\overline{D_2}}$ has parameters [1120,8,729] and weight enumerator $1 + 2240z^{729} + 4320z^{756}$.
\end{ex}

\begin{thm} \label{th6}
Let $C_{D_2}$ be defined by \eqref{cdd}, where $D_2$ is defined in \eqref{d2}.
If $\frac{m}{v} \equiv 0 \pmod{4}$, then $C_{D_2}$ is a $[p^{2m-1}+p^{m+v}-p^{m+v-1}-1, 2m]$ three-weight linear code with weight distribution given in Table \ref{tt4}.
\end{thm}

\begin{table} 
\caption{The weight distribution of $C_{D_2}$ if $\frac{m}{v} \equiv 0 \pmod{4}$}
\centering
\begin{tabular}{ll} 
\hline
Weight $w$                & Multiplicity $A_w$ \\
\hline
0 &1 \\
$(p-1)p^{2m-2}$ & $(p^{m-v}-1)(p^{m-v-1}+1)$ \\
$(p-1)p^{2m-2}(1+\frac{p-1}{p^{m-v}})$ &  $p^{2m}-p^{2m-2v}$\\
$(p-1)p^{2m-2}(1+\frac{1}{p^{m-v-1}})$ &  $(p^{m-v}-1)p^{m-v-1}(p-1)$ \\
\hline
\end{tabular}
 \label{tt4}
\end{table}

\begin{ex}
Let $p=3$, $m=4$ and $u=3$. 
Then $v=1$, $\frac{m}{v}=4$ and $s=2$.
The code $C_{D_2}$ has parameters [2348,8,1458] and weight enumerator $1 + 260z^{1458} + 5832z^{1566} + 468z^{1620}$. The code $C_{\overline{D_2}}$ has parameters [1174,8,729] and weight enumerator $1 + 260z^{729} + 5832z^{783} + 468z^{810}$.
\end{ex}

\section{The proofs of main results} \label{wt}

\subsection{Some auxiliary results}
\begin{lem}\label{l9}
Let
\[
n_1=|D_1|=|\{ (x,y) \in \mathbb{F}_q^2 \backslash \{(0,0)\}: \text{Tr}(x+y^{p^u+1})=0\}|.
\]
Then $n_1=p^{2m-1}-1$.
\end{lem}

\begin{pf}
By the orthogonal property of additive characters
\begin{align*}
n_1 &=\sum_{x,y \in \mathbb{F}_q} \frac{1}{p} \sum_{z \in \mathbb{F}_p} \zeta_p^{z \text{Tr}(x+y^{p^u+1})}-1 \\
&=\frac{q^2}{p}-1+ \frac{1}{p}\sum_{z \in \mathbb{F}_p^*} \sum_{x \in \mathbb{F}_q}\zeta_p^{\text{Tr}(zx)}S_u(z,0) \\
&=p^{2m-1}-1.
\end{align*}
\end{pf}

\begin{lem} \label{p}
If $\frac{m}{v}$ is even, then for $x \in \mathbb{F}_p^*$, 
\[
x^{\frac{q-1}{p^v+1}}=1.
\]
\end{lem}

\begin{pf}
Consider 
\[
\frac{q-1}{(p^v+1)(p-1)}=\frac{(p^s-1)(p^s+1)}{(p^v+1)(p-1)}.
\]
If $\frac{s}{v}$ is odd, then $(p^v+1)(p-1) \mid (p^s-1)(p^s+1)$ for $p^v+1 \mid p^s+1$ and $p-1 \mid p^s-1$.
If $\frac{s}{v}$ is even, it's easy to see $gcd(p-1, p^v+1)=2$.
So $(p^v+1)(p-1) \mid (p^s-1)(p^s+1)$ for $p^v+1 \mid p^s-1$ and $p-1 \mid p^s-1$.
\end{pf}

\begin{lem}\label{l10}
Let
\[
n_2=|D_2|=|\{ (x,y) \in \mathbb{F}_q^2 \backslash \{(0,0)\}: \text{Tr}(x^2+y^{p^u+1})=0\}|.
\]
Then if $\frac{m}{v}$ is odd or $\frac{m}{v} \equiv 2 \pmod{4}$,
\[
n_2=\begin{cases}
(p^m+1)(p^{m-1}-1), &  \text{if $p \equiv 3 \pmod{4}$ and $v$ is odd}, \\
(p^m-1)(p^{m-1}+1), & \text{otherwise}. 
\end{cases} 
\]
If $\frac{m}{v} \equiv 0 \pmod{4}$, 
\[
n_2=p^{2m-1}+p^{m+v}-p^{m+v-1}-1.
\]
Note that if $p \equiv 3 \pmod{4}$ and $m=1$, $D_2=\{ (x,y) \in \mathbb{F}_p^2 \backslash \{(0,0)\}: x^2+y^2=0 \}=\varnothing$, so $n_2=0$.
\end{lem}

\begin{pf}
By the orthogonal property of additive characters
\begin{align}
n_2 &=\sum_{x,y \in \mathbb{F}_q} \frac{1}{p} \sum_{z \in \mathbb{F}_p} \zeta_p^{z \text{Tr}(x^2+y^{p^u+1})}-1 \notag \\
&=\frac{q^2}{p}-1+ \frac{1}{p}\sum_{z \in \mathbb{F}_p^*}Q(z,0) S_u(z,0) \notag \\
&=p^{2m-1}-1+\frac{1}{p} \Omega,
\label{c0}
\end{align}
where $\Omega=\sum_{z \in \mathbb{F}_p^*}Q(z,0) S_u(z,0)$.
Then we evaluate $\Omega$ through three cases.

\begin{enumerate}
\item If $\frac{m}{v}$ is odd, by Lemma \ref{ga}
\[
G(\eta)^2=(p^*)^{m}=\begin{cases}
-p^m, &  \text{if $p \equiv 3 \pmod{4}$ and $v$ is odd}, \\
p^m, & \text{otherwise}. 
\end{cases}
\]
By Lemma \ref{s3} and \ref{qu} 
\begin{align}
\Omega &=(p-1)G(\eta)^2 \notag \\
&=\begin{cases}
-(p-1)p^m, &  \text{if $p \equiv 3 \pmod{4}$ and $v$ is odd}, \\
(p-1)p^m, & \text{otherwise}. 
\end{cases}
\label{c34}
\end{align}

\item If $\frac{m}{v} \equiv 2 \pmod{4}$, by Lemma  \ref{ga}
\[
G(\eta)=- (p^*)^{s}=\begin{cases}
p^s, &  \text{if $p \equiv 3 \pmod{4}$ and $v$ is odd}, \\
-p^s, & \text{otherwise}. 
\end{cases}
\]
By Lemma \ref{e}, \ref{s4}, \ref{qu} and \ref{p}
\begin{align}
\Omega &=-p^sG(\eta) \sum_{z_1 \in \mathbb{F}_p^*}\eta(z_1) \notag \\
&=\begin{cases}
-(p-1)p^m, &  \text{if $p \equiv 3 \pmod{4}$ and $v$ is odd}, \\
(p-1)p^m, & \text{otherwise}. 
\end{cases}
\label{c35}
\end{align}

\item If $\frac{m}{v} \equiv 0 \pmod{4}$,  by Lemma  \ref{ga} $G(\eta)=- p^s$ and by Lemma \ref{e}, \ref{s4}, \ref{qu} and \ref{p}
\begin{equation}
\Omega=-p^{s+v}G(\eta) \sum_{z_1 \in \mathbb{F}_p^*}\eta(z_1) =(p-1)p^{m+v}. 
\label{c36}
\end{equation}
\end{enumerate}
By \eqref{c0},\eqref{c34},\eqref{c35} and \eqref{c36} we complete the proof.
\end{pf}

\begin{lem} \label{a1}
If $m$ is odd, then
\[
B_1=|\{x \in \mathbb{F}_q: \text{Tr}(x^{p^u+1})=0 \}|=p^{m-1}.
\]
\end{lem}

\begin{pf}
By Lemma \ref{e} and \ref{s3}
\begin{align*}
B_1 &=\sum_{x \in \mathbb{F}_q} \frac{1}{p} \sum_{z \in \mathbb{F}_p} \zeta_p^{z \text{Tr}(x^{p^u+1})} \\
&=\frac{1}{p}\sum_{x \in \mathbb{F}_q}\left( 1+\sum_{z \in \mathbb{F}_p^*} \zeta_p^{z \text{Tr}(x^{p^u+1})} \right) \\
&=p^{m-1}+\frac{1}{p} \sum_{z \in \mathbb{F}_p^*} S_u(z,0) \\
&=p^{m-1}+\frac{G(\eta)}{p} \sum_{z \in \mathbb{F}_p^*} \eta_p(z) \\
&=p^{m-1}. 
\end{align*}
\end{pf}

\begin{lem} \label{a2}
If $\frac{m}{v} \equiv 0 \pmod{4}$, then
\[
B_2=|\{c \in \mathbb{F}_q: X^{p^{2u}}+X=c^{p^u} \ \text{is solvable in $\mathbb{F}_q$} \}|=p^{m-2v}.
\]
\end{lem}

\begin{pf}
The map 
\[
x \mapsto x^{p^{2u}}+x, \ x \in \mathbb{F}_q
\]
is $\mathbb{F}_p$-linear.
By Lemma \ref{sol}, the equation $X^{p^{2u}}+X=0$ has $p^{2v}$ solutions in $\mathbb{F}_q$.
Note that  $X^{p^u}$ is a permutation polynomial over $\mathbb{F}_q$, so $B_2=p^{m-2v}$.
\end{pf}

\subsection{The proofs of Theorems \ref{th1}, \ref{th2}, \ref{th3} and \ref{th4}}
By Lemma \ref{l9} the code $C_{D_1}$ has length $n_1=p^{2m-1}-1$.
For a codeword $c(a,b)$, $(a,b) \in \mathbb{F}_q^2 \backslash \{(0,0)\}$,
we will show that the Hamming weight $W_H(c(a,b))>0$, so the dimension of $C_{D_1}$ is $2m$.

We begin with the following equation:
\begin{equation*}
X^{p^{2u}}+X=(a^{-1}b)^{p^u}. \tag{E1}
\label{ab}
\end{equation*}
By Lemma \ref{sol},  it's not always solvable over $\mathbb{F}_q$ if $\frac{m}{v} \equiv 0 \pmod{4}$, and has an unique solution otherwise.  
Let $\gamma_{a,b}$  be some solution of \eqref{ab} if it exists.
Note that for $a, z_2 \in \mathbb{F}_p^*$ the equation $(-z_2a)^{p^u}X^{p^{2u}}+(-z_2a)X=-(z_2b)^{p^u}$ is equivalent to \eqref{ab}.
For $(a,b) \in \mathbb{F}_q^2 \backslash \{(0,0)\}$, we consider
\begin{align}
N_1(a,b) &=|\{(x,y) \in \mathbb{F}_q^2 \backslash \{(0,0)\}: \text{Tr}(x+y^{p^u+1})=0 \ and \ \text{Tr}(ax+by)=0\}| \notag \\ 
&=\sum_{x,y \in \mathbb{F}_q} \left( \frac{1}{p} \sum_{z_1 \in \mathbb{F}_p} \zeta_p^{z_1\text{Tr}(x+y^{p^u+1})} \right)  \left( \frac{1}{p} \sum_{z_2 \in \mathbb{F}_p} \zeta_p^{z_2 \text{Tr}(ax+by)} \right)-1 \notag\\
&=\frac{1}{p^2} \sum_{x,y \in \mathbb{F}_q} \left(1+ \sum_{z_1 \in \mathbb{F}_p^*} \zeta_p^{z_1\text{Tr}(x+y^{p^u+1})} \right)  \left(1+ \sum_{z_2 \in \mathbb{F}_p^*}\zeta_p^{z_2 \text{Tr}(ax+by)} \right)-1 \notag \\
&=p^{2m-2}-1+\frac{1}{p^2}\left(\sum_{z_1 \in \mathbb{F}_p^*}\sum_{x \in \mathbb{F}_q}\zeta_p^{\text{Tr}(z_1x)} S_u(z_1,0)+ \right. \notag \\
&\quad \left. \sum_{z_2 \in \mathbb{F}_p^*}\sum_{x \in \mathbb{F}_q}\zeta_p^{ \text{Tr}(z_2ax)}\sum_{y \in \mathbb{F}_q}\zeta_p^{ \text{Tr}(z_2by)}+\sum_{z_1, z_2 \in \mathbb{F}_p^*} \sum_{x \in \mathbb{F}_q} \zeta_p^{\text{Tr}((z_1+z_2a)x)} S_u(z_1,z_2b) \right) \notag \\
&=\begin{cases}
p^{2m-2}-1, &  \text{if}\ a \notin \mathbb{F}_p^*, \\
p^{2m-2}-1+p^{m-2} \Omega, & \text{if}\  a \in \mathbb{F}_p^*,
\end{cases}
\label{c2}
\end{align}
where 
\[
\Omega=\sum_{z_2 \in \mathbb{F}_p^*}S_u(-z_2a, z_2b),
\]
and it's clear that
\begin{equation} \label{wh}
W_H(c(a,b))=n_1-N_1(a,b).
\end{equation} 
 
If $m$ is odd, by Lemma \ref{e} and \ref{s1}
\begin{align} 
\Omega&=G(\eta) \sum_{z_2 \in \mathbb{F}_p^*}\eta(-z_2a) \bar{\chi} \left( -z_2a \gamma_{a,b}^{p^u+1}\right)  \notag \\
&=G(\eta) \sum_{z_2 \in \mathbb{F}_p^*}\eta_p(-z_2a) \zeta_p^{z_2a \text{Tr}(\gamma_{a,b}^{p^u+1})} \notag \\
&=\begin{cases}
0, &  \text{if}\  \text{Tr}(\gamma_{a,b}^{p^u+1})=0, \\
G(\eta)G(\eta_p)\eta_p \left( -\text{Tr}(\gamma_{a,b}^{p^u+1}) \right), & \text{if}\   \text{Tr}(\gamma_{a,b}^{p^u+1}) \ne 0.
\end{cases} \label{c61}
\end{align}
By Lemma \ref{ga}
\begin{equation} \label{cg}
G(\eta)G(\eta_p)= (p^*)^{\frac{m+1}{2}}.
\end{equation}
By \eqref{c2},\eqref{wh},\eqref{c61} and \eqref{cg} we know that the three nonzero weights of $C_{D_1}$ are $w_1=(p-1)p^{2m-2}$, $w_2=(p-1)p^{2m-2}\left( 1-\frac{1}{(p-1)p^{\frac{m-1}{2}}} \right)$, $w_3=(p-1)p^{2m-2}\left( 1+\frac{1}{(p-1)p^{\frac{m-1}{2}}} \right)$ and by Lemma \ref{a1}, $A_{w_2}+A_{w_3}=|\{(a,b) \in  \mathbb{F}_q^2: a \in \mathbb{F}_p^* \text{ and Tr}(\gamma_{a,b}^{p^u+1}) \ne 0\}|=(p-1)(q-p^{m-1})$.

If $\frac{m}{v}$ is odd and $v$ is even, by Lemma \ref{e} and \ref{s1}
\begin{align} 
\Omega&=G(\eta) \sum_{z_2 \in \mathbb{F}_p^*}\eta(-z_2a) \bar{\chi} \left( -z_2a \gamma_{a,b}^{p^u+1}\right)  \notag \\
&=G(\eta) \sum_{z_2 \in \mathbb{F}_p^*} \zeta_p^{z_2a \text{Tr}(\gamma_{a,b}^{p^u+1})} \notag \\
&=\begin{cases}
(p-1)G(\eta), &  \text{if}\  \text{Tr}(\gamma_{a,b}^{p^u+1})=0, \\
-G(\eta), & \text{if}\   \text{Tr}(\gamma_{a,b}^{p^u+1}) \ne 0.
\end{cases} \label{c62}
\end{align}
By Lemma \ref{ga}, $|G(\eta)|=p^s$.
By \eqref{c2},\eqref{wh} and \eqref{c62} we know that the three nonzero weights of $C_{D_1}$ are $w_1=(p-1)p^{2m-2}$, $w_2=(p-1)p^{2m-2}\left( 1-\frac{G(\eta)}{q} \right)$, $w_3=(p-1)p^{2m-2}\left( 1+\frac{G(\eta)}{(p-1)q} \right)$ and $A_{w_2}+A_{w_3}=|\{(a,b) \in  \mathbb{F}_q^2: a \in \mathbb{F}_p^*\}|=(p-1)q$.

If $\frac{m}{v} \equiv 2 \pmod{4}$, by Lemma \ref{s1}
\begin{align}
\Omega &=-p^s \sum_{z_2 \in \mathbb{F}_p^*}\bar{\chi} \left( -z_2a \gamma_{a,b}^{p^u+1}\right) \notag \\
&=\begin{cases}
-(p-1) p^s, &  \text{if}\  \text{Tr}(\gamma_{a,b}^{p^u+1})=0, \\
p^s, & \text{if}\   \text{Tr}(\gamma_{a,b}^{p^u+1}) \ne 0.
\end{cases}  \label{c63}
\end{align}
By \eqref{c2},\eqref{wh} and \eqref{c63} we know that the three nonzero weights of $C_{D_1}$ are $w_1=(p-1)p^{2m-2}$, $w_2=(p-1)p^{2m-2}\left( 1+\frac{1}{p^s} \right)$, $w_3=(p-1)p^{2m-2}\left( 1-\frac{1}{(p-1)p^s} \right)$ and $A_{w_2}+A_{w_3}=|\{(a,b) \in  \mathbb{F}_q^2: a \in \mathbb{F}_p^*\}|=(p-1)q$.

If $\frac{m}{v} \equiv 0 \pmod{4}$, by Lemma \ref{s2}
\begin{equation}
\Omega=0
\label{c64}
\end{equation}
or (\eqref{ab} is solvable)
\begin{align} 
\Omega &=-p^{s+v} \sum_{z_2 \in \mathbb{F}_p^*}\bar{\chi} \left( -z_2a \gamma_{a,b}^{p^u+1}\right) \notag \\
&=\begin{cases}
-(p-1)p^{s+v}, &  \text{if Tr}(\gamma_{a,b}^{p^u+1})=0, \\
p^{s+v}, & \text{if Tr}(\gamma_{a,b}^{p^u+1}) \ne 0.
\end{cases} \label{c65}
\end{align}
By \eqref{c2},\eqref{wh},\eqref{c64} and \eqref{c65} we know that the three nonzero weights of $C_{D_1}$ are $w_1=(p-1)p^{2m-2}$, $w_2=(p-1)p^{2m-2}\left( 1+\frac{1}{p^{s-v}} \right)$, $w_3=(p-1)p^{2m-2}\left( 1-\frac{1}{(p-1)p^{s-v}} \right)$ and by Lemma \ref{a2}, $A_{w_2}+A_{w_3}=|\{(a,b) \in  \mathbb{F}_q^2: a \in \mathbb{F}_p^* \text{ and  \eqref{ab} is solvable over $\mathbb{F}_q$} \}|=p^{m-2v}(p-1)$.

The weight distribution of $C_{D_1}$ follows from the result of $A_{w_2}+A_{w_3}$ and the first two Pless power moments.

\subsection{The proofs of Theorems \ref{th5} and \ref{th6}}
The proofs are similar to previous theorems.
The length of $C_{D_2}$ is given in Lemma \ref{l10}. 
For $(a,b) \in \mathbb{F}_q^2 \backslash \{(0,0)\}$,
we consider 
\begin{align}
N_2(a,b) &=|\{(x,y) \in \mathbb{F}_q^2 \backslash \{(0,0)\}: \text{Tr}(x^2+y^{p^u+1})=0 \ and \ \text{Tr}(ax+by)=0\}| \notag \\
&=\sum_{x,y \in \mathbb{F}_q} \left( \frac{1}{p} \sum_{z_1 \in \mathbb{F}_p} \zeta_p^{z_1\text{Tr}(x^2+y^{p^u+1})} \right)  \left( \frac{1}{p} \sum_{z_2 \in \mathbb{F}_p} \zeta_p^{z_2 \text{Tr}(ax+by)} \right)-1 \notag\\
&=p^{2m-2}-1+\frac{1}{p^2}(\Omega_1+\Omega_2),  
\label{c1}
\end{align}
where
\begin{equation} 
\Omega_1=\sum_{z_1 \in \mathbb{F}_p^*}Q(z_1,0) S_u(z_1,0)
\label{c3}
\end{equation}
and 
\begin{equation} 
\Omega_2=\sum_{z_1, z_2 \in \mathbb{F}_p^*} Q(z_1,z_2a) S_u(z_1,z_2b).
\label{c4}
\end{equation}
Note that we have discussed $\Omega_1$ in Lemma \ref{l10}.
Let $\gamma_b \in \mathbb{F}_q$ be some solution of the equation
\begin{equation*}
X^{p^{2u}}+X=-b^{p^u} \tag{E2}
\label{cd}
\end{equation*}
if it exists, then for $z_1, z_2 \in \mathbb{F}_p^*$, $z_3 \gamma_b$ is the solution of $z_1^{p^u}X^{p^{2u}}+z_1 X=-(z_2 b)^{p^u}$, where $z_3=z_1^{-1} z_2$.

By Lemma \ref{e}, \ref{s1} and \ref{qu}, if $\frac{m}{v}$ is odd,
\begin{align} 
\Omega_2&=G(\eta)^2 \sum_{z_1, z_3 \in \mathbb{F}_p^*} \bar{\chi }\left(\frac{z_1 z_3^2 a^2}{4}+z_1 (z_3 \gamma_b)^{p^u+1}\right)  \notag \\
&=G(\eta)^2 \sum_{z_1, z_3 \in \mathbb{F}_p^*} \zeta_p^{-z_1 z_3^2 \text{Tr}\left(\frac{a^2}{4}+ \gamma_b^{p^u+1}\right)} \notag \\
&=\begin{cases}
(p-1)^2G(\eta)^2, &  \text{if}\  \text{Tr}\left(\frac{a^2}{4}+ \gamma_b^{p^u+1}\right)=0, \\
-(p-1)G(\eta)^2, & \text{if}\   \text{Tr}\left(\frac{a^2}{4}+ \gamma_b^{p^u+1}\right) \ne 0.
\end{cases} \label{c66}
\end{align}
And if $\frac{m}{v} \equiv 2 \pmod{4}$,
\begin{align}
\Omega_2 &=-p^sG(\eta) \sum_{z_1, z_3 \in \mathbb{F}_p^*}\eta(z_1) \bar{\chi }\left(\frac{z_1 z_3^2 a^2}{4}+z_1 (z_3 \gamma_b)^{p^u+1}\right)  \notag \\
&=-p^sG(\eta) \sum_{z_1, z_3 \in \mathbb{F}_p^*} \zeta_p^{-z_1 z_3^2 \text{Tr}\left(\frac{a^2}{4}+ \gamma_b^{p^u+1}\right)} \notag \\
&=\begin{cases}
-(p-1)^2p^sG(\eta), &  \text{if}\  \text{Tr}\left(\frac{a^2}{4}+ \gamma_b^{p^u+1}\right)=0, \\
(p-1)p^sG(\eta), & \text{if}\   \text{Tr}\left(\frac{a^2}{4}+ \gamma_b^{p^u+1}\right) \ne 0.
\end{cases} 
\label{c67}
\end{align}
From the proof of Lemma \ref{l10} and \eqref{c1},\eqref{c3},\eqref{c4},\eqref{c66},\eqref{c67} we know that the two nonzero weights of $C_{D_2}$ are $w_1=(p-1)p^{2m-2}$ and
\[
w_2=\begin{cases}
(p-1)p^{2m-2}(1-\frac{1}{p^{m-1}}), &  \text{if $p \equiv 3 \pmod{4}$ and $v$ is odd}, \\
(p-1)p^{2m-2}(1+\frac{1}{p^{m-1}}), & \text{otherwise}.
\end{cases}
\]

If $\frac{m}{v} \equiv 0 \pmod{4}$, by Lemma \ref{ga}, \ref{e}, \ref{s2} and \ref{qu}
\begin{equation}
\Omega_2=0
\label{c68}
\end{equation}
or (\eqref{cd} is solvable)
\begin{align}
\Omega_2 &=-p^{s+v}G(\eta) \sum_{z_1, z_3 \in \mathbb{F}_p^*}\eta(z_1) \bar{\chi }\left(\frac{z_1 z_3^2 a^2}{4}+z_1 (z_3 \gamma_b)^{p^u+1}\right)  \notag \\
=& p^{m+v} \sum_{z_1, z_3 \in \mathbb{F}_p^*} \zeta_p^{-z_1 z_3^2 \text{Tr}\left(\frac{a^2}{4}+ \gamma_b^{p^u+1}\right)}  \notag \\
=& \begin{cases}
(p-1)^2p^{m+v}, &  \text{if}\  \text{Tr}\left(\frac{a^2}{4}+ \gamma_b^{p^u+1}\right)=0, \\
-(p-1)p^{m+v}, & \text{if}\   \text{Tr}\left(\frac{a^2}{4}+ \gamma_b^{p^u+1}\right) \ne 0.
\end{cases} 
\label{c69}
\end{align}
By \eqref{c1},\eqref{c3},\eqref{c4},\eqref{c36},\eqref{c68} and \eqref{c69} we know that the three nonzero weights of $C_{D_2}$ are $w_1=(p-1)p^{2m-2}$, $w_2=(p-1)p^{2m-2}(1+\frac{p-1}{p^{m-v}})$, $w_3=(p-1)p^{2m-2}(1+\frac{1}{p^{m-v-1}})$ and by Lemma \ref{a2}, $A_{w_2}=|\{(a,b) \in  \mathbb{F}_q^2 \backslash \{(0,0)\}: \eqref{cd} \text{ has no solution over } \mathbb{F}_q \}|=q(q-p^{m-2v})$.

The weight distribution of $C_{D_2}$ follows from the first two Pless power moments.

\section{Conclusions} \label{clu}
In this paper, inspired by the work in \cite{li2016construction}, several classes of two-weight and three-weight linear codes were constructed with their weight distributions settled using Weil sums.  
Some optimal or almost optimal linear codes were found.
It would be interesting if more linear codes with few weights can be presented.


\bibliographystyle{plain} 
\bibliography{thesis}
\end{document}